\documentclass[prl,aps,superscriptaddress,twocolumn]{revtex4-1}
\usepackage[utf8]{inputenc}
\usepackage[T1]{fontenc}
\usepackage[english]{babel} 
\usepackage[colorlinks=true,linkcolor=blue,anchorcolor=blue,citecolor=blue,urlcolor=blue]{hyperref}
\usepackage{mathrsfs,amssymb,amsthm,amsmath}
\usepackage{graphicx,subfigure,epsfig,epsf}
\usepackage{appendix,multirow}
\graphicspath{{figures/}}
\usepackage[usenames]{color}
\definecolor{Red}{rgb}{1,0,0}
\definecolor{Blue}{rgb}{0,0,1}
\newcommand{\bra}[1]{\langle #1|}
\newcommand{\ket}[1]{|#1\rangle}

\begin{document}

\title{Stable States with Non-Zero Entropy under Broken $\mathcal{PT}$-Symmetry}

\author{Jingwei Wen}
\affiliation{State Key Laboratory of Low-Dimensional Quantum Physics and Department of Physics, Tsinghua University, Beijing 100084, China}

\author{Chao Zheng}
\affiliation{Department of Physics, College of Science, North China University of Technology - Beijing 100144, China}

\author{Zhangdong Ye}
\affiliation{State Key Laboratory of Low-Dimensional Quantum Physics and Department of Physics, Tsinghua University, Beijing 100084, China}

\author{Tao Xin}
\email{xint@sustech.edu.cn}
\affiliation{Shenzhen Institute for Quantum Science and Engineering, Southern University of Science and Technology, Shenzhen 518055, China}
\affiliation{Guangdong Provincial Key Laboratory of Quantum Science and Engineering, Southern University of Science and Technology, Shenzhen 518055, Guangdong, China}

\author{Guilu Long}
\email{gllong@tsinghua.edu.cn}
\affiliation{State Key Laboratory of Low-Dimensional Quantum Physics and Department of Physics, Tsinghua University, Beijing 100084, China}
\affiliation{Frontier Science Center for Quantum Information, Beijing 100084, China}
\affiliation{Beijing National Research Center for Information Science and Technology, Beijing 100084, China}
\affiliation{Beijing Academy of Quantum Information Sciences, Beijing 100193, China}

\begin{abstract}

The $\mathcal{PT}$-symmetric non-Hermitian systems have been widely studied and explored both in theory and in experiment these years due to various interesting features. In this work, we focus on the dynamical features of a triple-qubit system, one of which evolves under local $\mathcal{PT}$-symmetric Hamiltonian. A new kind of abnormal dynamic pattern in the entropy evolution process is identified, which presents a parameter-dependent stable state, determined by the non-Hermiticity of Hamiltonian in the broken phase of $\mathcal{PT}$-symmetry. The entanglement and mutual information of a two-body subsystem can increase beyond the initial values, which do not exist in the Hermitian and two-qubit $\mathcal{PT}$-symmetric systems. Moreover, an experimental demonstration of the stable states in non-Hermitian system with non-zero entropy and entanglement is realized on a four-qubit quantum simulator with nuclear spins. Our work reveals the distinctive dynamic features in the triple-qubit $\mathcal{PT}$-symmetric system and paves the way for practical quantum simulation of multi-party non-Hermitian system on quantum computers.
 
\end{abstract}

\maketitle
\emph{Introduction.---}In the conventional quantum mechanics, the Hamiltonian of a closed system requires to be Hermitian \cite{Nielsen}, which guarantees the reality of the energy spectrum and the unitarity of the corresponding time evolution operators. However, the Hermiticity requirement is a sufficient condition but not necessary for real eigenvalues, and in 1998 \cite{Bender}, Bender and Boettcher found that a class of Hamiltonians satisfying joint $\mathcal{P}$ (spatial reflection) and $\mathcal{T}$ (time reversal) symmetry instead of Hermiticity can still have real eigenvalues in the unbroken phase \cite{nonlinear,add_pt}. Moreover, there exists critical points for phase transition from the $\mathcal{PT}$ unbroken phase to broken phase, called exceptional point or branch point \cite{EP1,cir_EP,pt_the1}. Because of various peculiar characters in this kind of non-Hermitian system, such as the violation of no-signaling principle \cite{violation,violation_exp}, entanglement restoration \cite{entanglepra2014, wenpt} and reversible-irreversible criticality in information flow \cite{flow2017theory,flow2019exp}, the $\mathcal{PT}$-symmetric quantum mechanics has aroused continuous attention and research interest in many perspectives. Recently, there are some related researches on its potential applications in reconstructing standard quantum theory \cite{reconstruct,reconstruct2} and it has been shown that a unitary evolution can be introduced by redefining the inner product of quantum states  \cite{pt_inner2015,Nori_nogo}, which make it equivalent to the Hermitian quantum theory.

\begin{figure}
 	 \centering
	     \includegraphics[scale=0.28]{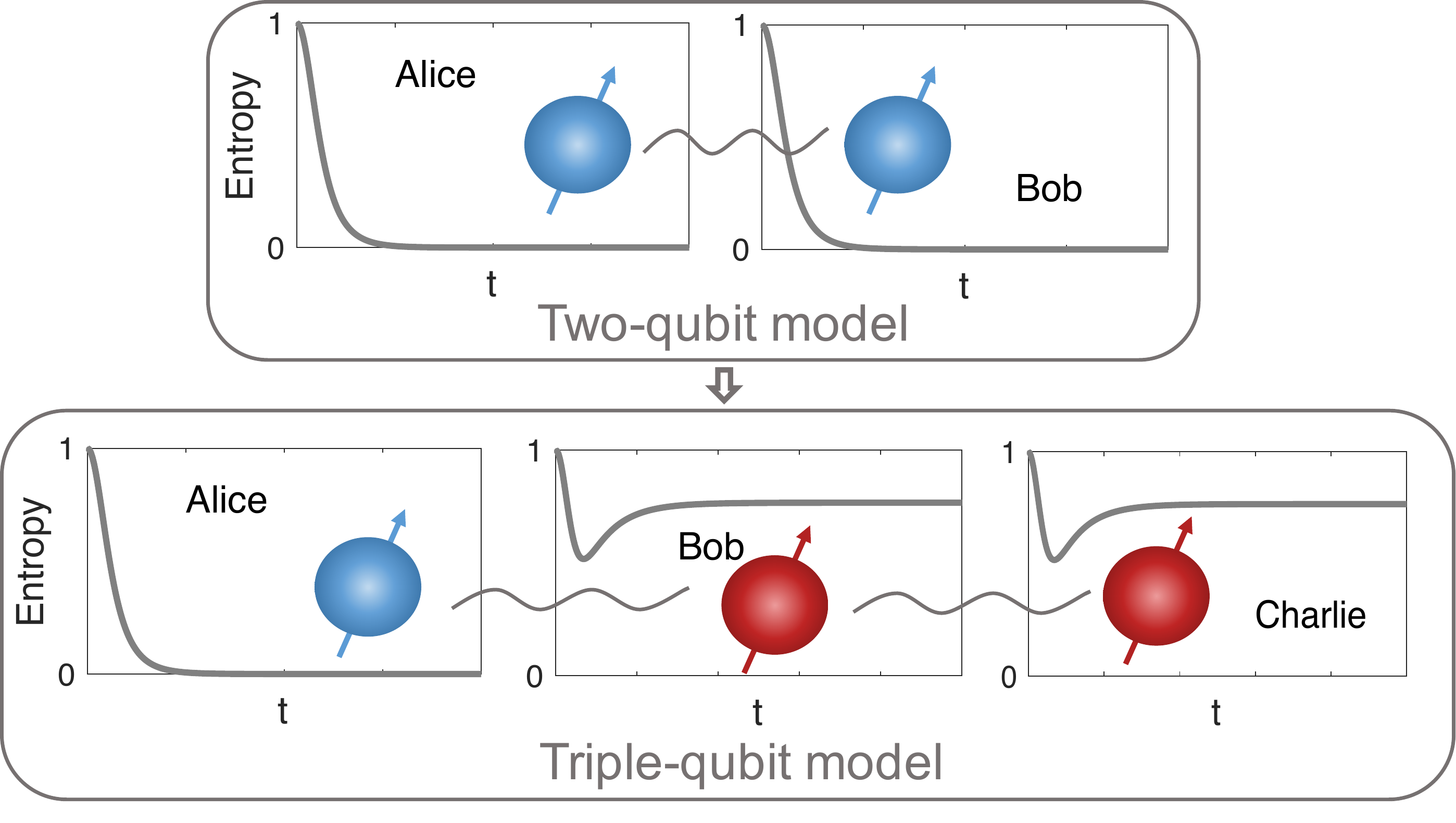}
          \caption{The multi-qubit system with local $\mathcal{PT}$-symmetric operators. Quantum system is initialized as maximally entangled state and the first qubit (Alice) undergoes a local $\mathcal{PT}$-symmetric operation, while the other qubits remain isolated. The black lines are entropy evolution with time in broken $\mathcal{PT}$-symmetric phase.}
          \label{model} 
\end{figure}

In experiment, many quantum processes such as symmetry-breaking transitions \cite{exp_transition1,exp_transition2,exp_transition3,exp_transition4,exp_transition5}, observation of exceptional point \cite{exp_ep1,exp_ep2} and topological features \cite{exp_topo1,exp_topo2,exp_topo3} of the $\mathcal{PT}$-symmetric system have been demonstrated and it depends mainly on the optical systems \cite{flow2019exp}, nuclear spins \cite{wenpt,exp_pt0}, ultracold atoms \cite{exp_transition1}, nitrogen-vacancy centers \cite{exp_transition2}, and superconductor systems \cite{exp_ep1} by introducing balanced gain and loss or state-selective dissipation. Moreover, some previous researches \cite{violation_exp,wenpt,flow2019exp} focus on the two-body non-Hermitian system as shown in Fig. \ref{model}, where two qubits (Alice and Bob) are entangled initially and one of them (Alice) evolves under local $\mathcal{PT}$-symmetric Hamiltonian. Such a two-qubit model can lead to oscillations of entropy and entanglement in the unbroken phase of $\mathcal{PT}$-symmetry, which violates the property of entanglement monotonicity \cite{entanglepra2014, wenpt}. Especially, the entropy and entanglement of both qubits will decay exponentially to zero in the broken phase and form stable states, which do not vary with time. Such stable states, whose dynamic process is named normal dynamic pattern (NDP) here, is just related to the quantum phase but independent on the degree of non-Hermiticity.

However, in this work, we find that when the system is extended from two-body to triple-body model, another kind of evolution process, named abnormal dynamic pattern (ADP) here, arises up. Subsystems evolving under ADP can present novel non-Hermiticity-related stable states with non-zero entropy. By controlling the local system of Alice, the entanglement and mutual information between Bob and Charlie can be redistributed and even increased beyond the initial value, which do not exist in the two-qubit $\mathcal{PT}$-symmetric system. Some theoretical and numerical analyses are introduced to study the properties of the partial-information reserved quantum states in the broken phase of $\mathcal{PT}$-symmetry. By enlarging the system with ancillary qubits and encoding the subsystem with the non-Hermitian Hamiltonian with postselection, an experimental demonstration of the stable states in ADP is realized on a four-qubit quantum simulator based on quantum circuit algorithm. 

\emph{Entropy of stable states.---}We focus on the dynamical features of a composite system consisting of three qubits, which is initialized as Greenberger-Horne-Zeilinger (GHZ) state \cite{GHZ} $\ket{\psi_{0}}=(\ket{000}+\ket{111})/\sqrt{2}$ and the reduced density matrix of each single qubit is $\rho_{\textup{single}}=I/2$, which is the maximally mixed state. Then one of the qubits, such as Alice qubit, performs local operator $U_{A}=e^{-i\hat{H}_{\mathcal{PT}}t}$ (set $\hbar=1$) on her own system with $\mathcal{PT}$-symmetric Hamiltonian

\begin{equation}
\hat{H}_{\mathcal{PT}}=s(\sigma_{x}+ir \sigma_{z})
\label{eq1}
\end{equation}
where $\sigma_{i}~(i=x,y,z)$ are Pauli matrix. The parameter $s>0$ represents energy scale and $r>0$ is the degree of non-Hermiticity. The $\mathcal{PT}$-symmetric Hamiltonian $\hat{H}_{\mathcal{PT}}$ satisfies $(\mathcal{PT})\hat{H}_{\mathcal{PT}}(\mathcal{PT})^{-1}=\hat{H}_{\mathcal{PT}}$, where operator $\mathcal{P}=\sigma_{x}$ and $\mathcal{T}$ corresponds to complex conjugation. The energy gap of the Hamiltonian $w=2s\sqrt{1-r^{2}}$ will be real as long as $r<1$, which means the $\mathcal{PT}$-symmetry is unbroken. The condition $r>1$ will lead to a broken phase with a transition at exceptional point $r_{ep}=1$. The three-body Hamiltonian can be expressed as $\hat{H}_{\mathcal{PT}}^{3}=\hat{H}_{\mathcal{PT}}\otimes I_{B} \otimes I_{C}$. Density matrix $\rho(t)$ of the whole system can be obtained by time-evolving operator with renormalized quantum state \cite{flow2017theory}

\begin{equation}
\rho(t)=\frac{e^{-i\hat{H}_{\mathcal{PT}}^{3}t}\rho(0)e^{i\hat{H}_{\mathcal{PT}}^{3\dagger}t}}{\textup{tr}[e^{-i\hat{H}_{\mathcal{PT}}^{3}t}\rho(0)e^{i\hat{H}_{\mathcal{PT}}^{3\dagger}t}]} 
\label{evolution2}
\end{equation}

\begin{figure}
 	 \centering
	     \includegraphics[scale=0.34]{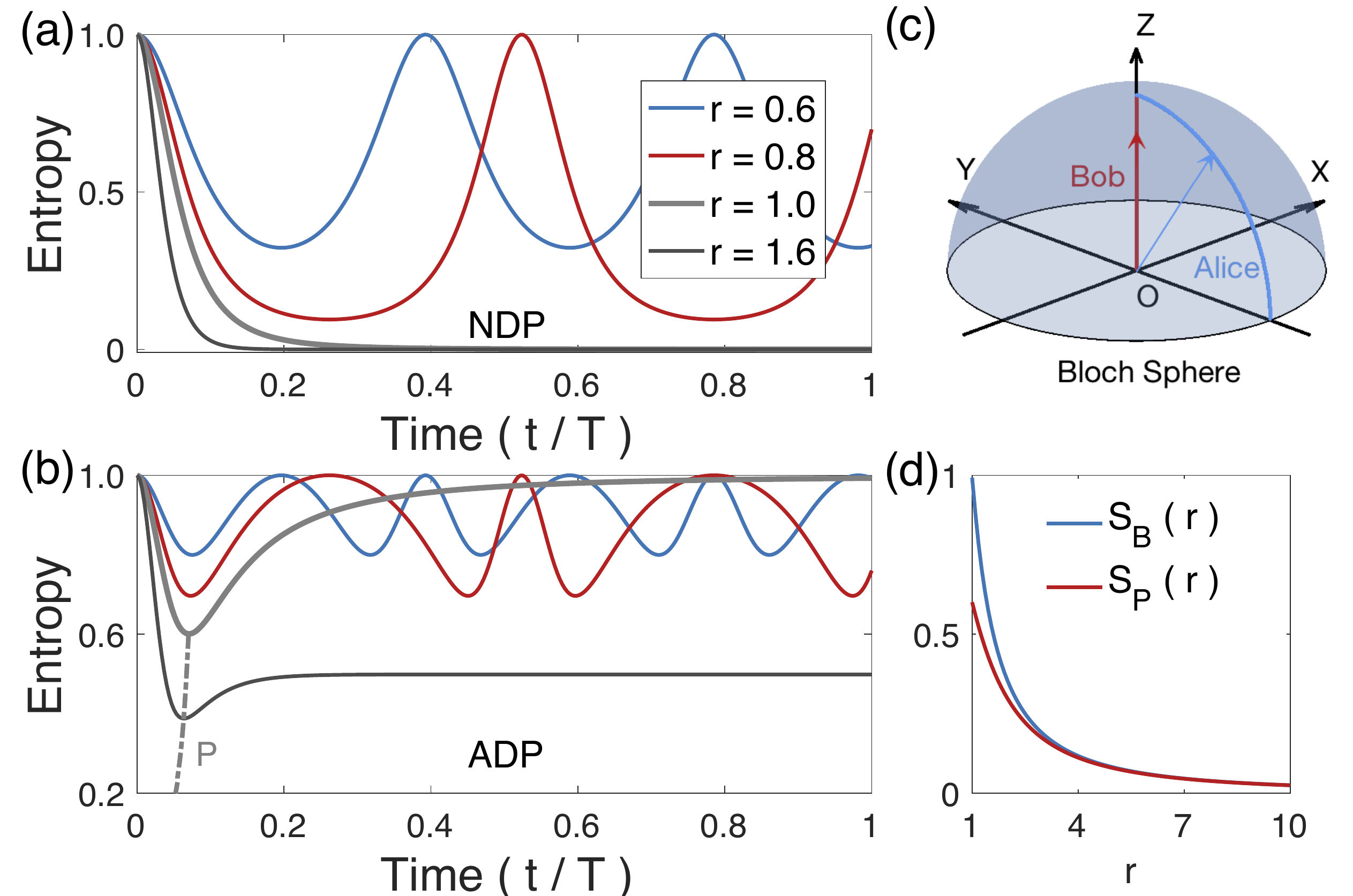}
          \caption{Two kinds of dynamical evolution pattern. (a) The entropy $S(\rho_{A})$ shows a NDP, while (b) $S(\rho_{B})$ has an ADP. (c) Bloch vectors of stable states labeled by lines with arrows in the Bloch upper hemisphere. Trajectory of Bloch vectors are represented by lines with corresponding colors when changing non-Hermiticity in broken phase. (d) The entropy at point $P$ and Bob's stable states. }
          \label{PT3_entropy} 
\end{figure}

\begin{figure*}
 	 \centering
          \includegraphics[scale=0.445]{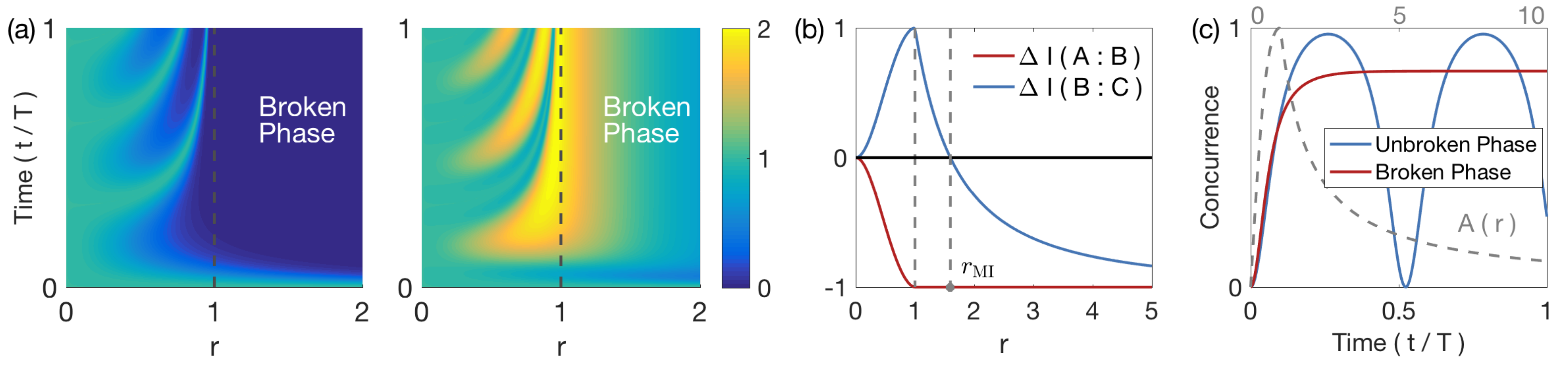} 
          \caption{(a) Mutual information of $\textup{I}(A:B)$ (left) and $\textup{I}(B:C)$ (right) with time under different non-Hermiticity.  (b) Increase of mutual information $\Delta \textup{I}(A:B)$ and $\Delta \textup{I}(B:C)$ of stable states with the degree of non-Hermiticity. (c) Concurrence of two-body subsystem (Bob-Charlie) in the different $\mathcal{PT}$-phases. The gray dashed line is the amplitude of concurrence with non-Hermiticity, while the other lines are concurrence evolution with time.}
          \label{fig_entanglement} 
\end{figure*}

The joint reduced states of two-body system are $\rho_{ij}=\textup{tr}_{k}(\rho)$, while the single-body reduced density matrices are $\rho_{i}=\textup{tr}_{jk}(\rho)~(i,j,k=A,B,C)$. We focus on the dynamical features of the von Neumann entropy $S(\rho)=-\textup{tr}(\rho \log_{2}\rho)$ \cite{Nielsen} and plot the evolution process within total time $T$ under different phases in Fig. \ref{PT3_entropy}\textcolor{blue}{(a), (b)}. It can be concluded that in the triple-qubit $\mathcal{PT}$-symmetric system, for the single-body subsystem, the entropy of Alice still evolve under NDP: entropy oscillates in the unbroken phase and the amplitude increases when the parameter $r$ approaching $r_{ep}$. Once crossing the exceptional point, entropy exponentially decays to zero and tend to be stable states, which are indistinguishable in terms of entropy evolution characteristics. However, the dynamic pattern of $S(\rho_{B})$ changed and another kind of ADP shows up:  entropy still oscillates in the unbroken phase, whereas in the broken phase of $\mathcal{PT}$-symmetry, the entropy of stable states will not decrease to zero exponentially, but stabilize to a value related to the degree of non-Hermiticity. In other words, there exist a parameter-dependent stable state in the subsystem of multi-party $\mathcal{PT}$-symmetric system and the entropy decreases with the increase of non-Hermiticity. Such a stable state can maintain partial-entropy in the system under broken $\mathcal{PT}$-symmetry.

Based on the evolution equation, we can determine the reduced density matrix of Bob qubit

\begin{equation}
\rho_{B}=\frac{1}{N}
\begin{smallmatrix}
\begin{pmatrix}
\vert C \vert^2+(A-B)^2&0\\
0&\vert C \vert^2+(A+B)^2 \\ 
\end{pmatrix}
\end{smallmatrix}
\end{equation}
where $A=\cos(wt/2)$, $B=(-2rs/w)\sin(wt/2)$, $C=(-2is/w)\sin(wt/2)$ and $N=2(\vert C \vert^2+A^2+B^2)$ is the normalization constant. We focus on the stable state in broken phase with $\rho_{B}^{\textup{ss}}=\frac{1}{2}I+\frac{\sqrt{r^2-1}}{2r}\sigma_{z}$. At exceptional point, the density matrix of stable state is a maximally mixed state with entropy $S(\rho_{B})=1$. However, with the increase of non-Hermiticity, the stable state will tend to be $\rho_{B}^{\textup{ss}}=\ket{0}\bra{0}$, a pure state with entropy $S(\rho_{B})=0$. It can serve as a quantum state purification phenomenon induced by the non-Hermiticity increase and the pure state is a stable state with time. Then we can calculate the analytical expression of the von Neumann entropy for Bob qubit in ADP

\begin{equation}
\begin{split}
S(\rho_{B}^{\textup{ss}})=\log_{2}\frac{2}{\cos\theta(\sec\theta+\tan\theta)^{\sin\theta}}>0
\end{split}
\label{Sb}
\end{equation}
where $\cos\theta=1/r$ and $\theta \in [0,\pi/2)$ \cite{supp}. The entropy will not be zero unless $\theta=\pi/2$, which means the non-Hermiticity of system is infinity. As for the entropy evolution in NDP, the density matrix of Alice in stable state with broken $\mathcal{PT}$-symmetry is $\rho_{A}^{\textup{ss}}=\rho_{B}^{\textup{ss}}-D(r)\cdot\sigma_{y}$. It can be found that the stable states of Alice and Bob have same population distribution, but the $\rho_{A}^{\textup{ss}}$ has off-diagonal elements, which decrease in power-law with a damping function $D(r)=1/(2r)$ \cite{supp}. Such effects can be modeled as a phase damping process induced by non-Hermiticity, leading to the loss of quantum information to the environment. As shown in Fig. \ref{PT3_entropy}\textcolor{blue}{(c)}, we plot the Bloch vectors of stable states of Alice and Bob in Bloch sphere. When $r$ increases from the exceptional point to a large enough value, the Bloch vector of Alice's stable state rotates along the Bloch sphere surface from point $(0,-1,0)$ towards the north point of $z$-axis with norm $\vert \vert \vec{r}_{A} \vert \vert = 1$ all the process. Therefore, the entropy  of the stable states is

\begin{equation}
S(\rho_{A}^{\textup{ss}})=-\sum_{i=1,2}\lambda_{i}^{A}\log_2{\lambda_{i}^{A}} \equiv 0
\end{equation}
with eigenvalues $\lambda_{1,2}^{A}=0,1$, which are not related to the non-Hermiticity parameter and this is what happens in the evolution process obeying NDP. However, the norm of the Bloch vector of Bob in stable states is $\vert \vert \vec{r}_{B} \vert \vert =\sin\theta\le 1$, which starts at the center of Bloch sphere at exceptional point and move towards the top point when increasing non-Hermiticity. Moreover, besides the stable states, there exist another kind of non-Hermiticity-related quantum states satisfying $dS(t)/dt=0$ at the specific points $P=(t_{P}(r),S_{P}(r))$ labeled by gray dashed line in Fig. \ref{PT3_entropy}\textcolor{blue}{(b)}. During the evolution from point $P$ to stable states in ADP, the entropy increase and this turning point does not exist in NDP. With the increase of non-Hermiticity parameter, the entropy of quantum state at time point $t_{P}$ will gradually approach $S(\rho_{B}^{\textup{ss}})$ and the duration and intensity of the entropy increase process will gradually weaken until it disappears.

\emph{Entanglement evolution.---}Furthermore, we turn to investigate the dynamical features of interaction and entanglement in the triple-party $\mathcal{PT}$-symmetric system. Entropic quantities are generally used to quantify correlations and for a two-body system with density matrix $\rho_{ij}$, the amount of information shared between the two parts can be characterized by the mutual information defined as $\textup{I}(i:j)=S(\rho_i)+S(\rho_j)-S(\rho_{ij})\ge 0$. The mutual information is always non-negative, and cannot be zero unless $i$ and $j$ are in a separable state, ensuring that $\textup{I}(i:j)$ is a genuine measure of correlations \cite{mutual}. It is usually believed that local trace-preserving quantum operations can never increase mutual information \cite{Nielsen} but this can be violated in the two-qubit $\mathcal{PT}$-symmetric system without exceeding the initial value \cite{wenpt,flow2019exp}.

In the triple-qubit $\mathcal{PT}$-symmetric system, this property still can be hold in subsystem (Alice-Bob) under NDP, just as shown in Fig. \ref{fig_entanglement}\textcolor{blue}{(a)}. However, we find that evolution in the ADP of $\textup{I}(B:C)$, which has non-zero mutual information in broken phase, can present mutual information beyond the initial value. Moreover, the mutual information $\textup{I}(B:C)$ oscillates with a maximum deviation $\textup{I}_{\textup{md}}$ from the initial values at a series of discrete time points but tends to be stable at value $\textup{I}_{s}$ after passing exceptional point. We define a variation measure $\Delta \textup{I}(r)=\mathcal{O}(r_{ep}-r)\textup{I}_{\textup{md}}+\mathcal{O}(r-r_{ep})\textup{I}_{s}-\textup{I}(t_{0})$ to quantify the increase of mutual information, where $\mathcal{O}(\cdot)$ is the Heaviside step function. We can conclude from Fig. \ref{fig_entanglement}\textcolor{blue}{(b)} that the stable value $\textup{I}_{s}$ decreases with $r$ and the subsystem (Bob-Charlie) has the maximal available information at exceptional point. However, the increasing process of mutual information stops at $r_{\textup{MI}}\approx1.5978$, a critical point for increase of mutual information, which is different from the exceptional point of the $\mathcal{PT}$-symmetry. In other word, the critical point for transition of phase is not that for the increase of accessible information in triple-qubit $\mathcal{PT}$-symmetric system. It is noted that there exists an anti-corresponding relation between entropy and mutual information evolution because of the fact that $\textup{I}(B:C,t_\infty)=2S(\rho_B^{\textup{ss}})$ : the subsystems of two-body, which have NDP in entropy evolution, can present ADP in mutual information evolution, and vice verse.

To evaluate the degree of entanglement in the two-body subsystem, we can also use concurrence \cite{concurrence}
\begin{equation}
C(\rho_{ij})=\textup{Max} \{ 0,\sqrt{\lambda_{1}}-\sqrt{\lambda_{2}}-\sqrt{\lambda_{3}}-\sqrt{\lambda_{4}} \}
\end{equation}
where $\lambda_{i}$ is the eigenvalues of $\rho_{ij}(\sigma_{y}\sigma_{y})\rho^{*}_{ij}(\sigma_{y}\sigma_{y})$ in decreasing order. We numerically calculate the dynamical evolution and find that $C(\rho_{AB})=C(\rho_{AC})=0$ all the time, which do not show evolution. It is Alice's $\mathcal{PT}$-symmetric operators introduce the non-Hermiticity, but the systems including Alice qubit do not show entanglement oscillation and this is different from the two-qubit counterparts. But for $C(\rho_{BC})$, the concurrence will emerge both in the unbroken and broken phase, although the local Hamiltonian of them are Hermitian. This evolution pattern is consistent with the entropic quantities of mutual information. We identify the amplitude of concurrence during the evolution by $A(r)=C_{\textup{max}}$ and it can be concluded from Fig. \ref{fig_entanglement}\textcolor{blue}{(c)} that the concurrence of $\rho_{BC}^{\textup{ss}}$ will decrease and tend to be stable at $C_{s}=1/r$, which presents a power-law decay with the increase of non-Hermiticity after exceptional point.

\begin{figure}
 	 \centering
  	 \includegraphics[scale=0.85]{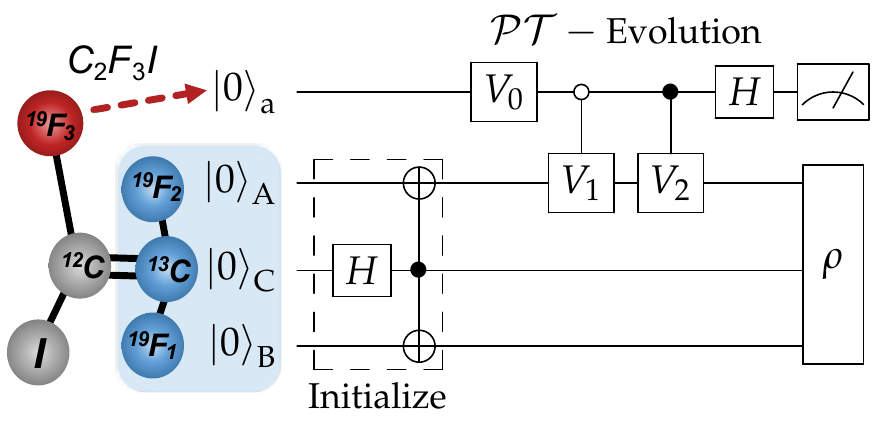}
          \caption{Experimental sample and quantum circuit. Three of the four controllable qubits are used as work system and the last one is an ancillary qubit. The whole process is divided into initial state preparation, $\mathcal{PT}$-symmetric evolution and measurement.}
          \label{fig_exp1} 
\end{figure}

\begin{figure}
 	 \centering
           \includegraphics[scale=0.43]{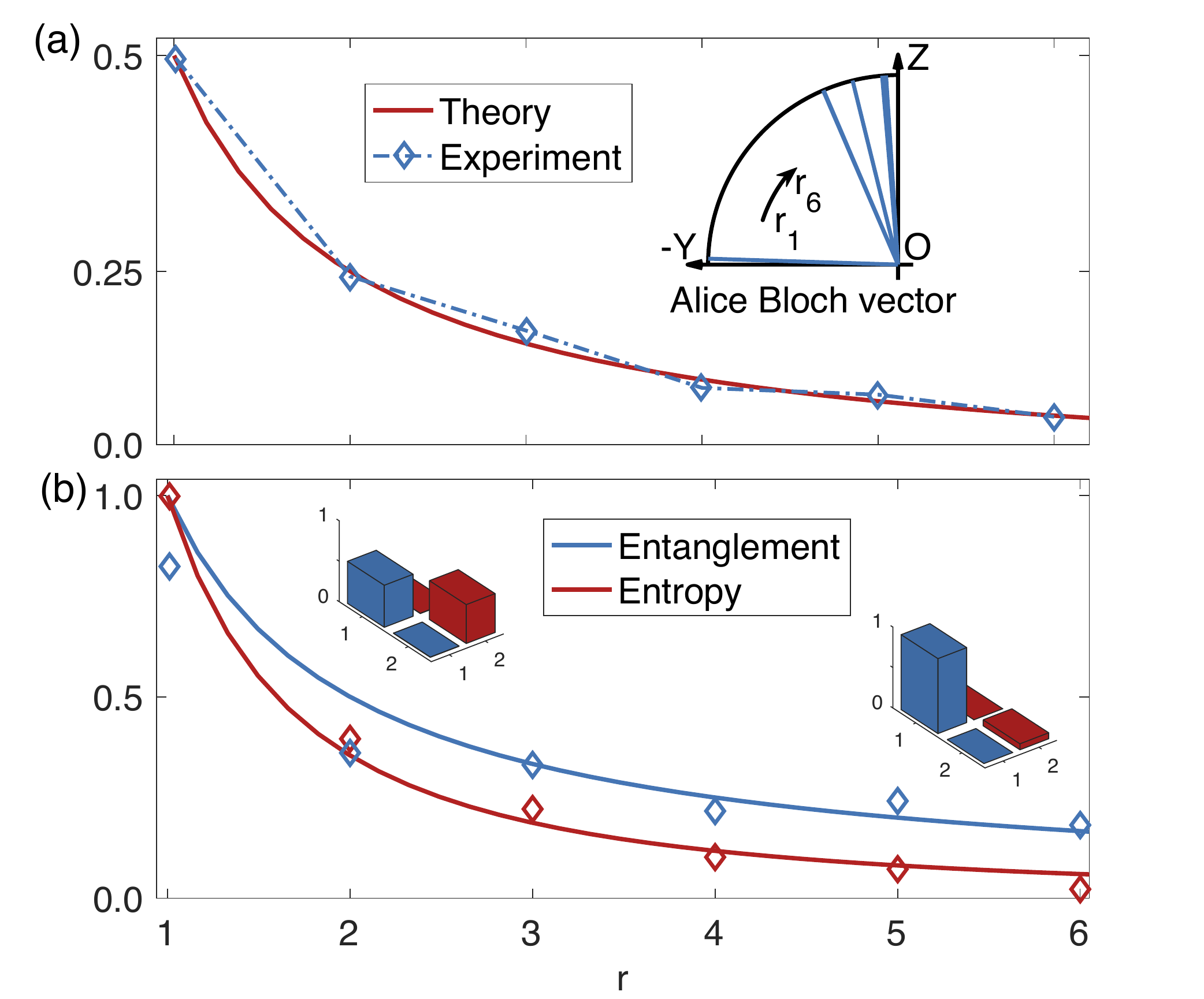}
          \caption{(a) The off-diagonal elements of stable states $\rho_{A}^{\textup{ss}}$ in NDP. The diamond points are experimental results while the solid line represents the theoretical expectations. The inset panel shows the experimentally identified Bloch vectors of Alice. (b) Experimental results of entropy and concurrence of the stable states in ADP. The inset panels represent the density matrix of Bob with minimal and maximal non-Hermiticity in experimental parameter setup.}
          \label{fig_exp2} 
\end{figure}

\emph{Experimental observation of stable states.---}In experiment, we focus on demonstrating the entropy dynamic evolution of stable states in the triple-qubit system with local $\mathcal{PT}$-symmetric operators on a liquid nuclear magnetic resonance quantum simulator. The sample used is ${}^{13}C$-labeled iodotrifluoroethylene (C$_{2}$F$_{3}$I) and the qubits in blue box of Fig. \ref{fig_exp1} encode the work system while another nucleus ${}^{19}F_{3}$ is chosen as ancillary system to realize $\mathcal{PT}$-symmetric operator \cite{wenpt}. The operators in the dotted box initialize the work system to the GHZ state. To realize the quantum simulation of the non-unitary evolution induced by $\mathcal{PT}$-symmetric Hamiltonian on Alice, we decompose the non-Hermitian Hamiltonian evolution into a linear combination of unitary operators and realize the simulation in an enlarged Hilbert space with post-selection \cite{dual1,dual2,dual3,dual4}. Notations $H$ in the quantum circuit represent Hardmard gate and the 1-controlled gate $V_{2}=\sigma_{z}$. The single-qubit operator $V_{0}$ and 0-controlled $V_{1}$ are parameter-dependent quantum gates and the concrete forms are

\begin{equation}
\begin{smallmatrix}
\begin{split}
V_{0}=\begin{pmatrix}
\cos\phi&-\sin\phi  \\
\sin\phi&\cos\phi
\end{pmatrix},~
V_{1}=\begin{pmatrix}
\cos\phi_{1}&i\sin\phi_{1} \\
i\sin\phi_{1}&\cos\phi_{1}
\end{pmatrix}
\end{split}
\end{smallmatrix}
\end{equation}
where $ \phi=\arcsin\frac{r\sin{(wt/2)}}{M_{1}}$, $\phi_{1}=\arcsin\frac{-\sin{(wt/2)}}{M_{2}}$ and $M_{1}=[1-r^2\cos{wt}]^{1/2}$, $M_{2}=[1-r^2\cos^2{(wt/2)}]^{1/2}$ \cite{supp}. Then the evolution can be realized via single-qubit operations and two-qubit controlled gates. We take several different parameter points in experiment and all the operations are realized using shaped pulses \cite{grape11,grape22}, while being robust to the static field distributions and inhomogeneity and the durations of the experimental pulses are within 15ms. At the end of quantum circuit, we obtain the density matrix of work system by observing the probe spin ${}^{13}C$ in the subspace $\ket{0}$ of ancillary qubit \cite{YangB}. We trace out different qubits of experimental stable states to find the subsystems with different dynamical patterns. 

As shown in Fig. \ref{fig_exp2}\textcolor{blue}{(a)}, the non-diagonal elements of the density matrix of Alice's stable states present a power-law decay with the increase of non-Hermiticity, which is consistent with the damping function $D(r)$. It is how quantum states behave in the NDP as analyzed above. In ADP, we experimentally determine the entropy of Bob and concurrence between Bob and Charlie in  Fig. \ref{fig_exp2}\textcolor{blue}{(b)}, which both present the parameter-related non-zero value in stable states with broken $\mathcal{PT}$-symmetry. The experimental results of entropy match well with the theoretical expectation of Eq. (\ref{Sb}) in different parameter conditions. The inset panels represent the density matrix of quantum states of Bob with minimal and maximal non-Hermiticity in experimental parameter setup with average fidelities over 0.989 and we can see that with the increase of non-Hermiticity, $\rho_{B}^{\textup{ss}}$ gradually evolves from the maximally mixed state to a pure state.

\emph{Conclusion.---}We investigate the evolution process of entropy and entanglement in a triple-qubit system with local $\mathcal{PT}$-symmetric operation from theoretical and experimental perspectives. Two kinds of dynamic pattern, named ADP and NDP, are found in this system, where entropy and entanglement tend to be stable at a non-Hermiticity-related non-zero value in the ADP, which do not exist in the two-qubit counterparts. Two-body subsystems in ADP present maximum entanglement increase at exceptional point and mutual information can increase beyond the initial values. A new critical point $r_{\textup{MI}}$ is determined in the broken phase, where the transition of accessible information from increase to decrease compared with the initial condition happens. Based on the four-qubit quantum simulator, we experimentally observe the stable states in non-Hermitian system with nuclear spins and the results confirmed the theoretical analysis. Our work shows that when the $\mathcal{PT}$-symmetric system is extended from two-body to triple-body, some different physical properties occur and the enhancement of entanglement and mutual information has important physical significance. Especially, there are some potential applications in quantum communication and quantum eavesdropping by regulating and controlling the channel capacity of system with local $\mathcal{PT}$-symmetric operators on the third party.

This work was supported by the National Key R$\&$D Program of China (2017YFA0303700), the Key R$\&$D Program of Guangdong province (2018B030325002), Beijing Advanced Innovation Center for Future Chip (ICFC) and the National Natural Science Foundation of China under Grants No. 11774197. C. Z is supported by National Natural Science Foundation of China Grant No. 11705004. T. X. is also supported by the National Natural Science Foundation of China (Grants No. 11905099, and No. U1801661), Guangdong Basic and Applied Basic Research Foundation (Grant No. 2019A1515011383), and Guangdong Provincial Key Laboratory (Grant No. 2019B121203002).


\clearpage
\section{Supplementary Material}

\section{Derivation of Entropy Evolution of Stable States}
In the triple-qubit system, the non-unitary operator on Alice induced by the $\mathcal{PT}$-symmetric Hamiltonian is

\begin{equation}
 \begin{smallmatrix}
\begin{split}
U_{A}=e^{i\phi}
\begin{pmatrix}
\cos\frac{wt}{2\hbar}+\frac{2rs}{w}\sin\frac{wt}{2\hbar}&\frac{-2is}{w}\sin\frac{wt}{2\hbar}\\
\frac{2is}{w}\sin\frac{wt}{2\hbar}&\cos\frac{wt}{2\hbar}-\frac{2rs}{w}\sin\frac{wt}{2\hbar}\\
\end{pmatrix}
\end{split}
\end{smallmatrix}
\end{equation}
where $\phi$ is a phase factor. Because the triple-qubit Hamiltonian is a direct product of each single-body Hamiltonian, the operator on the whole system can be expressed as $U_{3}=U_{A}\otimes I_{B} \otimes I_{C}$. Then the time-dependent quantum state of the triple-qubit system without considering the normalization constant is $\rho(t)=U_{3}\rho(0)U_{3}^{\dagger}$. We need to trace out the other qubits to find the density matrix of Bob, which presents an ADP in entropy evolution and it can be realized by

\begin{equation}
\begin{smallmatrix}
\begin{split}
\rho_{B}(t)=&\sum_{i,j=0,1}(\bra{i} \otimes I\otimes \bra{j})\rho(t)(\ket{j} \otimes I\otimes \ket{i})\\
=&\frac{1}{N}
\begin{pmatrix}
\vert C \vert^2+(A-B)^2&0\\
0&\vert C \vert^2+(A+B)^2 \\ 
\end{pmatrix}
\end{split}
\end{smallmatrix}
\end{equation}
where $A=\cos(wt/2\hbar)$, $B=(-2rs/w)\sin(wt/2\hbar)$, $C=(-2is/w)\sin(wt/2\hbar)$ and $N=2(\vert C \vert^2+A^2+B^2)$ is the normalization constant. In the unbroken phase of the $\mathcal{PT}$-symmetry, each item in the quantum state oscillates with time periodically. However, when the symmetry is broken, the energy gap will become a pure imaginary number and we set $w/2\hbar=ik$, where $k$ is a positive real number. According to the Euler equations, we can decompose each item in quantum state into exponentially increase item and exponentially decrease item, where the latter one can be abandoned in the long-time limit,

\begin{eqnarray}
\lim_{t \to \infty}
\begin{cases}
\cos^2\frac{wt}{2\hbar}=e^{2kt}/4 \\
\sin^2\frac{wt}{2\hbar}=-e^{2kt}/4\\
\cos^2\frac{wt}{2\hbar}\sin^2\frac{wt}{2\hbar}=ie^{2kt}/4\\
\end{cases}
\end{eqnarray}

Then the eigenvalues of the renormalized density matrix $\rho_{B}^{\textup{ss}}$ is $\lambda_{1,2}^{B}=\frac{r\pm\sqrt{r^2-1}}{2r}$ and we can calculate the analytical expression of the von Neumann entropy of stable state in ADP

\begin{equation}
\begin{split}
S(\rho_{B}^{\textup{ss}})=&-\sum_{i=1,2}\lambda_{i}^{B} \log_{2}\lambda_{i}^{B} \\
=& \log_{2}2r-\frac{\sqrt{r^2-1}}{r}\log_{2}(r+\sqrt{r^2-1}) \\
=& \log_{2}\frac{2}{\cos\theta(\sec\theta+\tan\theta)^{\sin\theta}} \\ 
\end{split}
\end{equation}
where $\cos\theta=1/r$ and $\theta \in [0,\pi/2)$. Based on the variable substitution, we can rewrite the quantum state as

\begin{equation}
\begin{split}
\rho_{B}^{\textup{ss}}=\frac{I+\sin\theta\cdot \sigma_{z}}{2}
\end{split}
\end{equation}
and the Bloch vector is $\vec{r}_{B}=(0,0,\sin\theta)$. So we can see that the purity of stable state is parameter-dependent and the stable state will evolve from a maximally mixed state to a pure state $\ket{0}$ when increasing non-Hermiticity. However, when we turn to the quantum stable state of Alice qubit, the density matrix have off-diagonal elements and the stable state in the broken phase is 

\begin{equation}
\begin{smallmatrix}
\begin{split}
\rho_{A}^{\textup{ss}}=&
\lim_{t\rightarrow\infty}
\frac{1}{N}
\begin{pmatrix}
\vert C \vert^2+(A-B)^2&2BC\\
-2BC&\vert C \vert^2+(A+B)^2 \\ 
\end{pmatrix} \\
=&
\begin{pmatrix}
\frac{r+\sqrt{r^2-1}}{2r} &\frac{-i}{2r}\\
\frac{i}{2r}&\frac{r-\sqrt{r^2-1}}{2r} \\ 
\end{pmatrix}\\
=&\frac{I-\cos\theta\cdot \sigma_{y}+\sin\theta\cdot \sigma_{z}}{2}
\end{split}
\end{smallmatrix}
\end{equation}
where the damping function $D(r)=\cos\theta/2=1/2r$ and the norm of the Bloch vector is $\vert \vert \vec{r}_{A} \vert \vert = \cos^2\theta+\sin^2\theta=1$, which is parameter-independent. The eigenvalues of the stable state of Alice is

\begin{equation}
\begin{split}
\lambda_{1,2}^{A}&=\frac{\rho_{A(11)}^{\textup{ss}}+\rho_{A(22)}^{\textup{ss}}}{2}\pm \\
&~~~\frac{\sqrt{(\rho_{A(11)}^{\textup{ss}}-\rho_{A(22)}^{\textup{ss}})^2+4\rho_{A(12)}^{\textup{ss}}\rho_{A(21)}^{\textup{ss}}}}{2}  \\
&=\frac{1\pm\sqrt{(r^2-1)/r^2+1/r^2}}{2}\\
&=0,1
\end{split}
\end{equation}
And this results that the entropy of stable states in the NDP keep zero all the time and is parameter-independent. This leads to that in the broken phase, the mutual information of subsystem Bob and Charlie is $I(B:C,t\to\infty)=S(\rho_B^{\textup{ss}})+S(\rho_C^{\textup{ss}})-S(\rho_{BC}^{\textup{ss}})=2S(\rho_B^{\textup{ss}})$. So by numerically solving the equation $S(\rho_B^{\textup{ss}})=1/2$, we can determine the critical point $r_{\textup{MI}}$ in the broken phase.

\section{Experimental Simulation of the Stable States}
   
\subsection{Initialization}
\begin{figure}
 	 \centering
  	\includegraphics[scale=0.6]{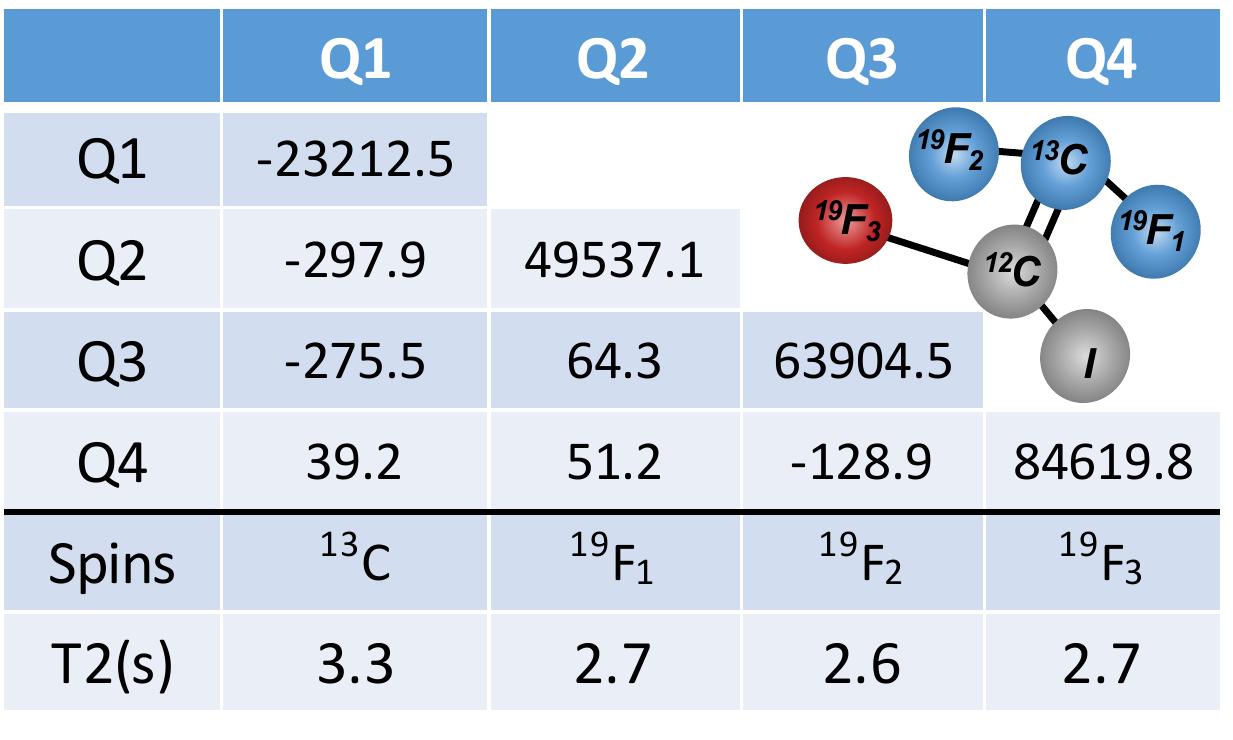}
          \caption{Molecule structure and molecule parameters of the sample.  ${}^{13}C$,  ${}^{19}F_{1}$,  ${}^{19}F_{2}$, and  ${}^{19}F_{3}$ are used as four qubits in experiment. The chemical shifts and J-couplings (in units of Hz) are listed by the diagonal and off-diagonal elements, respectively. The transversal relaxation time T2 (in seconds) are also shown at the bottom.}
          \label{fig_supp1} 
\end{figure}

The experiments for simulating the stable states in triple-qubit system with local $\mathcal{PT}$-symmetric operator are carried out on a 600 MHz nuclear magnetic resonance platform at room temperature (298 K) with a four-qubit sample ${}^{13}C$-labeled iodotrifluoroethylene dissolved in d chloroform. The spectrometer is equipped with a superconducting magnet which creates a strong magnetic field (14.1T). The sample is placed in the static magnetic field along the $z$-direction and the internal Hamiltonian under weak coupling approximation is 

\begin{equation}
\begin{split}
H_{int}=-\sum^{4}_{i=1}\pi\nu_{i}\sigma^{i}_{z}+\sum^{4}_{i<j}\frac{\pi}{2}J_{ij}\sigma^{i}_{z}\sigma^{j}_{z}
\end{split}
\end{equation}
where $\nu_{i}$ is the chemical shift and $J_{ij}$ is the J-coupling strength between the $i$th and $j$th nuclei. The experimentally identified parameters of this molecule is shown in Fig. \ref{fig_supp1}. The initialization process of quantum computation in liquid nuclear magnetic resonance system starts from a thermal equilibrium state obeying Boltzmann distribution:

\begin{equation}
\begin{split}
\rho_{eq}=\frac{e^{-H_{int}/k_{B}T}}{\textup{tr}(e^{-H_{int}/k_{B}T})}
\end{split}
\end{equation}
where $k_{B}$ is the Boltzmann constant and $T$ is the thermodynamic temperature. Under the condition that $\| H_{int}/k_{B}T\| \ll 1$ and $J_{kl}\ll \omega_{i}$, the thermal equilibrium state in our platform can be approximated as

\begin{equation}
\rho_{eq} \approx \frac{1}{2^{4}}(I^{\otimes 4}+\sum_{i}^{4}\frac{\hbar w_{i} \sigma_{z}^{i}}{2k_{B}T}) \label{eq2}
\end{equation} 
where the notation $I$ is identity matrix and $\sigma_{z}$ is a Pauli matrix. To initialize the system, we generally need to drive the quantum system from the highly mixed state $\rho_{eq}$, which can not be used as an initial state to the pseudo-pure state (PPS)

\begin{equation}
\ket{\rho_{\textup{pps}}} = \frac{1-\epsilon}{2^4}I^{\otimes 4} + \epsilon\ket{0000}\bra{0000}
\end{equation} 
where $\epsilon\approx10^{-5}$ is polarization. The first term can be neglected since the identity matrix does not evolve under any unitary propagator and cannot be observed. We prepared the PPS from the thermal equilibrium state with the selective-transition method \cite{pps_lineselect,Yang}, which is realized by unitary operators and field gradient pulses in the $z$-direction (Gz). The unitary operators redistribute the diagonal elements and the Gz pulse is used to eliminate the undesired coherence except the zero-quantum coherence of spins. After these processes, the PPS is prepared and this state serves as the starting point for subsequent computation tasks.

\subsection{Quantum Simulation of $\mathcal{PT}$-Symmetric Operator}
To realize the simulation of non-unitary dynamical process induced by $\mathcal{PT}$-symmetric Hamiltonian, we encode the non-unitary evolution into unitary process by adding ancillary qubit and form a gate-based quantum circuit, which is friendly for experiment. It is called as the linear combination of unitaries, which is a universal subroutine in designing and developing quantum algorithms \cite{dual11}. We first create superposition states on the ancillary system, and then perform controlled operations on the work system. The physical picture is that different unitary operations are implemented simultaneously on the work system but in different subspaces and the final results can be obtained in a specific subspace of ancillary systems according to the practical algorithm design. 

Specifically, suppose that the operator for creating superposition states is $V_{0}=[\cos\phi,-\sin\phi; \sin\phi,\cos\phi ]$ and the non-unitary evolution operator can be decomposed into the form $U_{A}=\cos\phi \cdot V_{1}+\sin\phi \cdot V_{2}$, where 

\begin{equation}
\begin{smallmatrix}
\begin{split}
V_{1}=\begin{pmatrix}
\cos\phi_{1}&i\sin\phi_{1} \\
i\sin\phi_{1}&\cos\phi_{1} \\
\end{pmatrix},
V_{2}=\begin{pmatrix}
\cos\phi_{2}&-i\sin\phi_{2}  \\
i\sin\phi_{2}&-\cos\phi_{2} \\
\end{pmatrix}
\end{split}
\end{smallmatrix}
\end{equation}

Under the unitary limitation on $V_{i}$ $(i=0,1,2)$, the choice of these operators are not unique. This construction leads to four equations as follows

\begin{eqnarray}
\begin{cases}
\cos\phi \cdot \cos\phi_{1}=\cos\frac{wt}{2\hbar}\\
\sin\phi \cdot \cos\phi_{2}=\frac{2rs}{w}\sin\frac{wt}{2\hbar}\\
\cos\phi \cdot \sin\phi_{1}=\frac{-2s}{w}\sin\frac{wt}{2\hbar}\\
\sin\phi \cdot \sin\phi_{2}=0\\
\end{cases}
\end{eqnarray}
 
\begin{figure}
 	 \centering
	\includegraphics[scale=0.9]{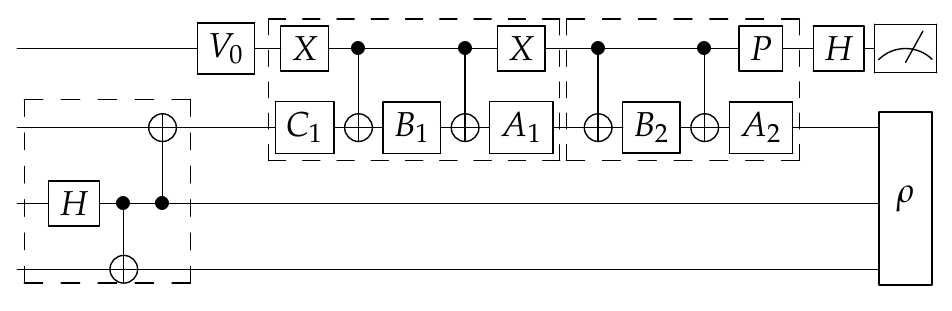}
          \caption{Quantum gate decomposition for the simulation of the triple-qubit system with local $\mathcal{PT}$-symmetric operator. The concrete forms of the single-qubit gates are shown in Table \ref{Tab:table}.}
          \label{fig_supp2} 
\end{figure}

By solving these equations, we can determine the angles $\phi$ and $\phi_{1,2}$ as shown in the main text. It worth noting that $\tan\phi_{2}=0$ and this leads to $V_{2}=\sigma_{z}$. Single-qubit operator $V_{0}$ and two-qubit operator $V_{1}$ are parameter-dependent quantum gates, while the other unitary quantum gates do not vary with the parameters in the $\mathcal{PT}$-symmetric Hamiltonian. In the broken phase of $\mathcal{PT}$-symmetry, the operators can be determined based on the experimental parameter setup and decomposed into single-qubit operations and controlled-NOT gates, as shown in Fig. \ref{fig_supp2}. Quantum evolution according to the quantum circuit we constructed are optimized by gradient ascent pulse engineering \cite{grape1,grape2}. Each shaped pulse is simulated to be over 99.5\% fidelity \cite{fidelity}, while being robust to the static field distributions and inhomogeneity and the durations of the experimental pulses are within 15ms. 

\begin{table}[htbp]
\caption{Decomposition scheme of quantum algorithm with single-qubit gates and controlled-NOT gate according to the parameter setting in experiment. The $R_{i}(\alpha)~(i=y,z)$ is a rotation operator along $i$-axis with angle $\alpha$.}
\centering
\renewcommand\arraystretch{1.5}
\begin{tabular}{cccccc}
\hline
\hline
$A_1$:&$R_{z}(1.5\pi)R_{y}(\phi_{1})$&$A_2$:&$R_{z}(1.5\pi)R_{y}(\pi)$&$V_{0}$:&$R_{y}(0.5\pi)$\\
$B_1$:&$R_{y}(-\phi_{1})R_{z}(-\pi)$&$B_2$:&$R_{y}(-\pi)R_{z}(-1.5\pi)$&P:&$\begin{pmatrix} \begin{smallmatrix} 1&0\\0&-i \end{smallmatrix} \end{pmatrix}$\\
$C_1$:&$R_{z}(-0.5\pi)$ &X:&\multicolumn{3}{c}{$R_{z}(-0.5\pi)R_{y}(\pi)R_{z}(0.5\pi)$}\\
\hline
\hline
\end{tabular}
\label{Tab:table}
\end{table}

\subsection{Measurement and Results}

\begin{figure}
 	 \centering
  	\includegraphics[scale=0.35]{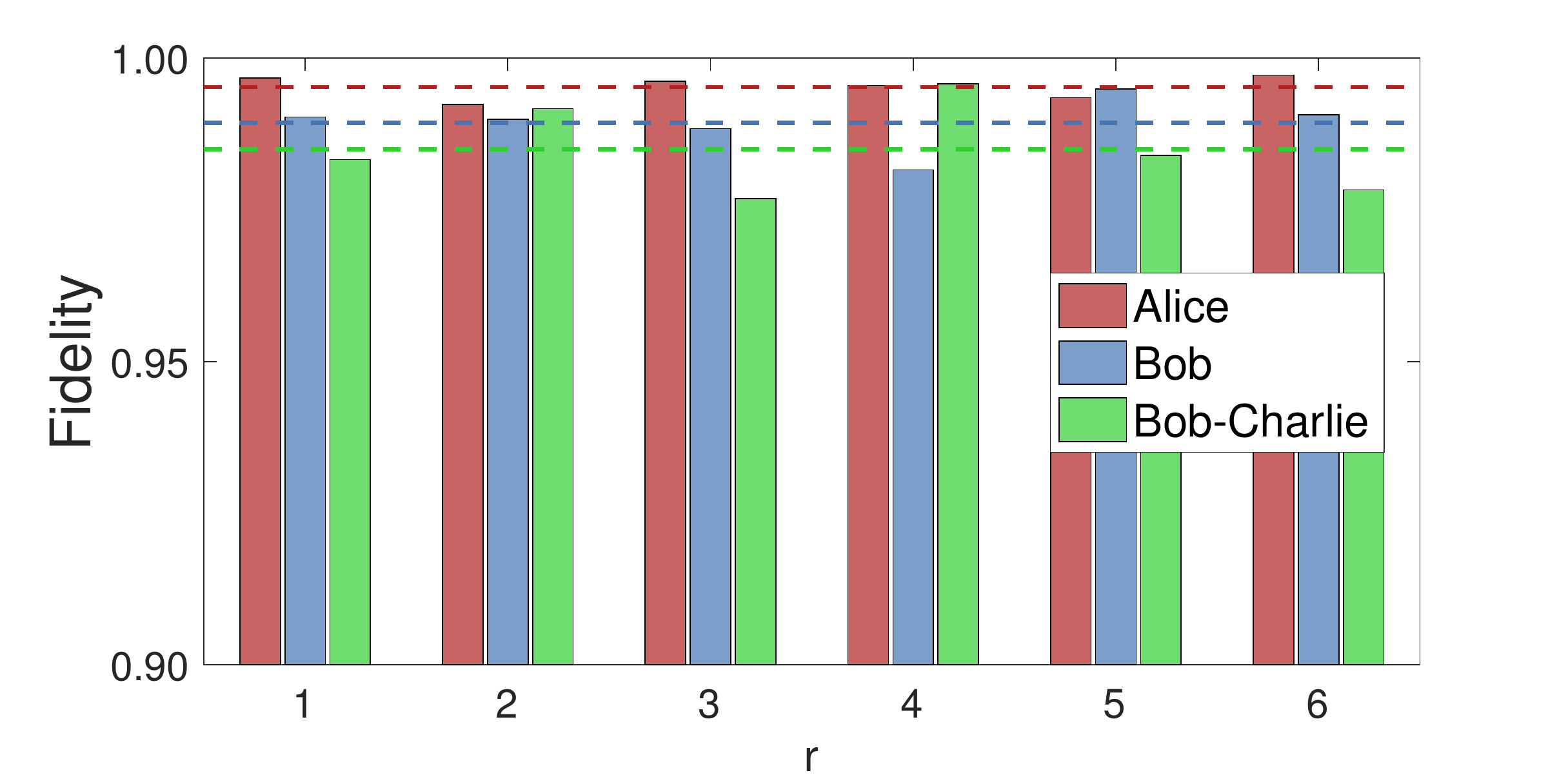}
          \caption{Fidelities of subsystems between the experimental stable states and the theoretical expectations under different non-Hermiticity. The average fidelities are labeled by lines with corresponding colors.}
          \label{fig_supp3} 
\end{figure}

\begin{figure}
 	 \centering
  	\includegraphics[scale=0.42]{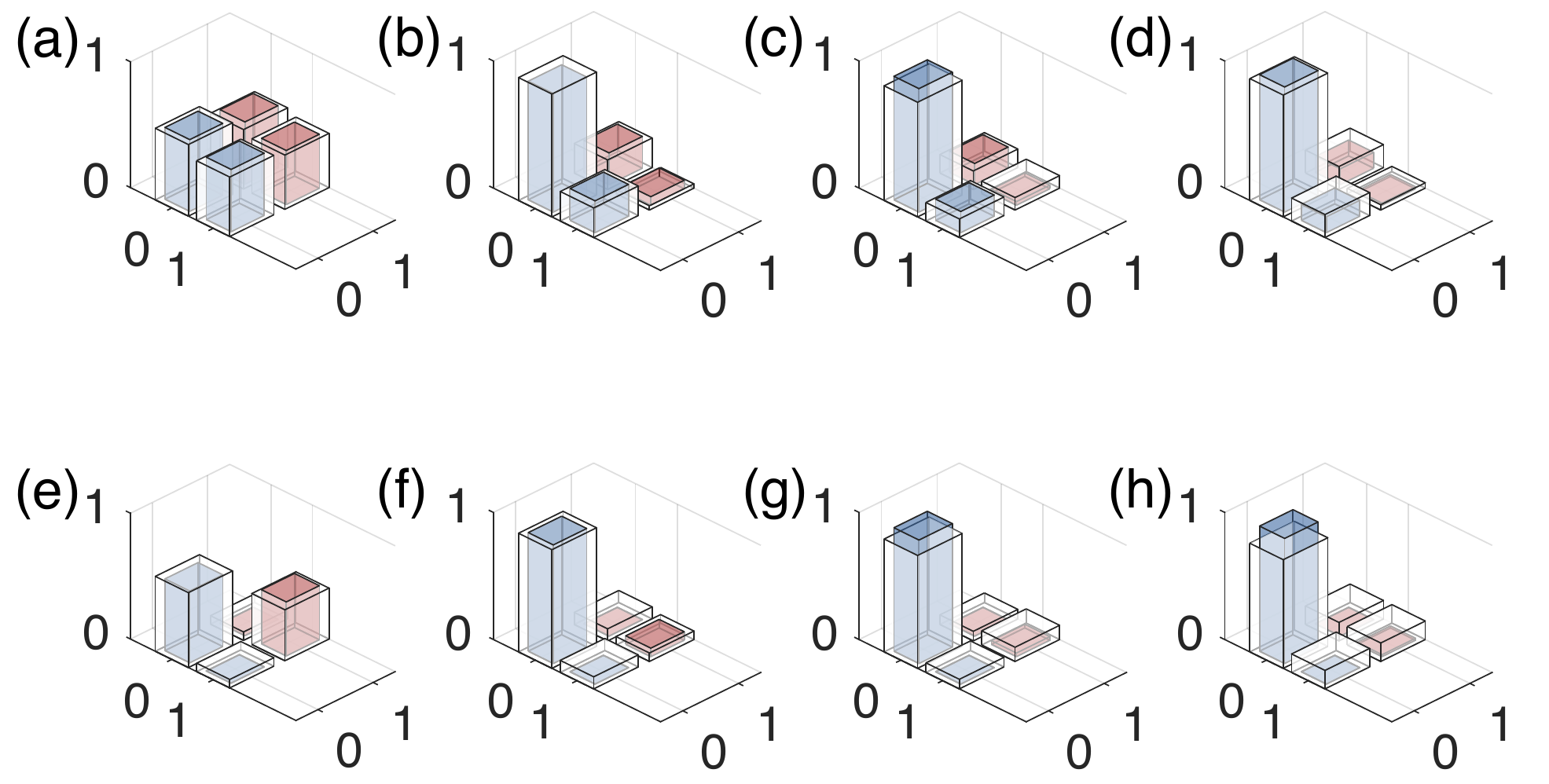}
          \caption{The experimentally identified density matrix of stable states under different degree of non-Hermiticity. Figures in the first row from (a) to (d) represent the quantum state of Alice in NDP ($r_1 \rightarrow r_{4}$), while figures in the second row from (e) to (h) show the density matrix of Bob in ADP. The external transparent bars represent the experimental results, while the internal solid bars represent the corresponding theoretical values of the density matrices. The quantum states are taken absolute value for better display.}
          \label{fig_supp4} 
\end{figure}

After the entanglement creation and $\mathcal{PT}$-symmetric evolution,  quantum measurement is performed on a bulk ensemble of molecules, which means the readout is an ensemble-averaged macroscopic measurement. At the end of the quantum circuit, all experimental data are extracted from the free-induction decay (FID), which is the signal induced by the precessing magnetization of the sample in a surrounding detection coil. The signal is then subjected to Fourier transformation, and the resulting spectral lines are fitted, yielding a set of measurement data. As the precession frequencies of different spins are distinguishable, they can be individually detected and all the observations are made on the probe spin ${}^{13}C$ \cite{Yang}. By fitting the ${}^{13}C$ spectrum, the real parts and the imaginary parts of the peaks are extracted, which correspond to $\langle \hat{\sigma}_{1}^{x} \rangle $ and $\langle\hat{\sigma}_{1}^{y}\rangle$, respectively. Then we can reconstruct all the density matrix elements in the subspace where the ancillary qubit is $\ket{0}$ to get the target stable states of the triple-qubit work system under different experimental parameter setup. We plot the fidelities of different subsystems between the experimental results and theoretical expectations in Fig. \ref{fig_supp3} with average fidelities over 0.98. The corresponding density matrix of stable states, which evolve under NDP and ADP respectively, are shown in Fig. \ref{fig_supp4} and both of them present quantum state purification phenomenon with the increase of non-Hermiticity.

\end{document}